\documentclass[acmsmall,nonacm, screen]{acmart}

\usepackage{graphicx}
\usepackage[many]{tcolorbox}
\usepackage{enumitem}
\usepackage{subcaption}
\usepackage[inkscapelatex=false]{svg}
\AtBeginDocument{%
  }

\begin{document}

% \title{Integrated sensing and communication: Perception, security, and trust in the 6G era}
\title{Future G Network's New Reality: Opportunities and Security Challenges}
\author{Chandra Thapa}
\email{chandra.thapa@data61.csiro.au}
\orcid{https://orcid.org/0000-0002-3855-3378}
\affiliation{%
  \institution{CSIRO Data61}
  \city{Sydney}
  \state{NSW}
  \country{Australia}
}

\author{Surya Nepal}
\email{surya.nepal@data61.csiro.au}
\orcid{https://orcid.org/0000-0002-3289-6599}
\affiliation{%
  \institution{CSIRO Data61}
  \city{Sydney}
  \state{NSW}
  \country{Australia}}

\renewcommand{\shortauthors}{Thapa et al.}

\begin{abstract}
Future G network's new reality is a widespread cyber-physical environment created by Integrated Sensing and Communication (ISAC). It is a crucial technology that transforms wireless connections into ubiquitous sensors. ISAC unlocks transformative new capabilities, powering autonomous systems, augmented human sensing, and next-generation immersive applications, such as digital twins. 
%
% Instead of just protecting data, the primary security concern now is maintaining the trustworthiness of perceptions, as the network's understanding of reality can be challenged by threats such as spoofing and jamming, which can cause physical harm. 
%
However, this new reality fundamentally reshapes the security landscape. 
%The primary security concern fundamentally shifts from merely safeguarding data to ensuring the trustworthiness of the system's perception of physical reality. 
%
The primary security concern shifts from the traditional focus on data protection to a new priority: safeguarding the integrity of the system’s perception of physical reality itself.
This perception can be perilously manipulated by sophisticated attacks such as sensing eavesdropping, phantom dangers, and invisible threats, potentially resulting in direct and catastrophic physical harm.
Traditional security measures, such as signature-based detection, are insufficient to counter these perception-level threats that mimic genuine physical signals. 
A proactive, layered, defense-in-depth strategy is required, integrating physical, environmental, intelligence, and architectural security measures to build a trustworthy ecosystem.
%
%Instead, a layered, defense-in-depth strategy is needed. This approach integrates physical, environmental, intelligence, and Zero-Trust security measures to foster a reliable ecosystem. 
%
Additionally, realizing ISAC's potential responsibly also depends on parallel efforts in global standardization and strong governance to address the significant challenges of privacy, liability, and the technology's dual-use.
%Additionally, establishing global standards and robust governance is crucial for effectively addressing privacy, liability, and dual-use concerns.
\end{abstract}
\maketitle

%====================================================
\section{Introduction}
% The future generation `\textbf{G}' of wireless communication networks is not just about speed; it is about giving our network eyes. Before we let it see, we must understand what it means for the network to perceive, decide what it is allowed to look at, and who is held responsible when its perception is deceived. 

% What is ISAC -- fundamental concept of ISAC with figure
%====================================================
%\subsection{The progression of wireless networks}
For decades, the story of each new wireless generation, each new `\textbf{G},' has focused on metrics such as speed, latency, and capacity~\cite{tariq}. 3G brought the internet to our pockets, 4G enabled the app economy and streaming videos, and 5G promised to connect our cities. The advent of Sixth-Generation (6G) systems, however, signifies a more substantial shift. The narrative is changing from merely enhancing communication to creating a cyber-physical continuum that seamlessly integrates the digital, physical, and human domains~\cite{ericsson_wp,saad}. 

\begin{figure}[th]
    \centering
    \includegraphics[trim={1cm 6cm 1cm 5.5cm},clip, width=0.98\linewidth]{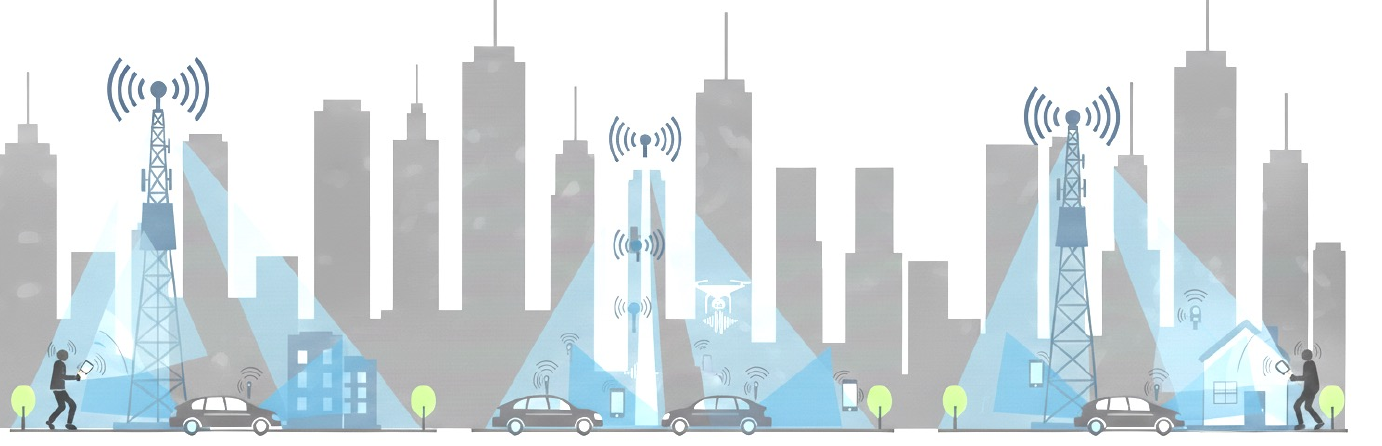}
    \caption{An illustration of Integrated Sensing and Communication (ISAC) with applications in a smart city setting and intelligent transportation. Communication towers transmit data and actively sense their environment using the same radio signals. The signals reflected off cars, drones, and people enable the network to determine their location, speed, and movement. The cars are sensing the objects and hazards and communicating with each other.}
    \label{fig:isac_city}
\end{figure}
At the heart of this vision is Integrated Sensing and Communication (ISAC)~\cite{huawei_wp,ISAC3}. ISAC unifies sensing and communication functionalities on a shared hardware and spectrum platform, a convergence driven by the technologies underpinning 6G, such as the move to higher-frequency bands and the use of massive antenna arrays~\cite{lima}. 
%
% ISAC offers benefits such as unprecedented gains in spectral and hardware efficiency, reduced hardware costs, and the ability to unlock new services by unifying sensing and communication functionalities. 
%
For the first time, the very same radio signals that provide our devices with a connection will also be used to perceive, map, and interact with the physical environment~\cite{huawei_wp,ISAC3}. 
%
%Its applications are vast and transformative, enabling high-accuracy localization, environmental mapping, augmented human sensing like vital sign monitoring, and powering critical infrastructure such as intelligent transportation systems, smart grids, and smart cities
%
Imagine if the air around us suddenly developed a form of echolocation, sensing the shape, location, and movement of objects and people. This is not a one-off experiment; the International Telecommunication Union (ITU), the UN agency that sets global telecommunication standards, has officially recognized ISAC as a core feature of 6G (IMT-2030)~\cite{5Gamerica_whitepaper,ITU_reco}.

\begin{figure}[t]
\begin{subfigure}[b]{0.5\textwidth} % Adjust width as needed
    \centering
    \includegraphics[trim={1cm 3cm 2.5cm 4.2cm},clip, width=\textwidth]{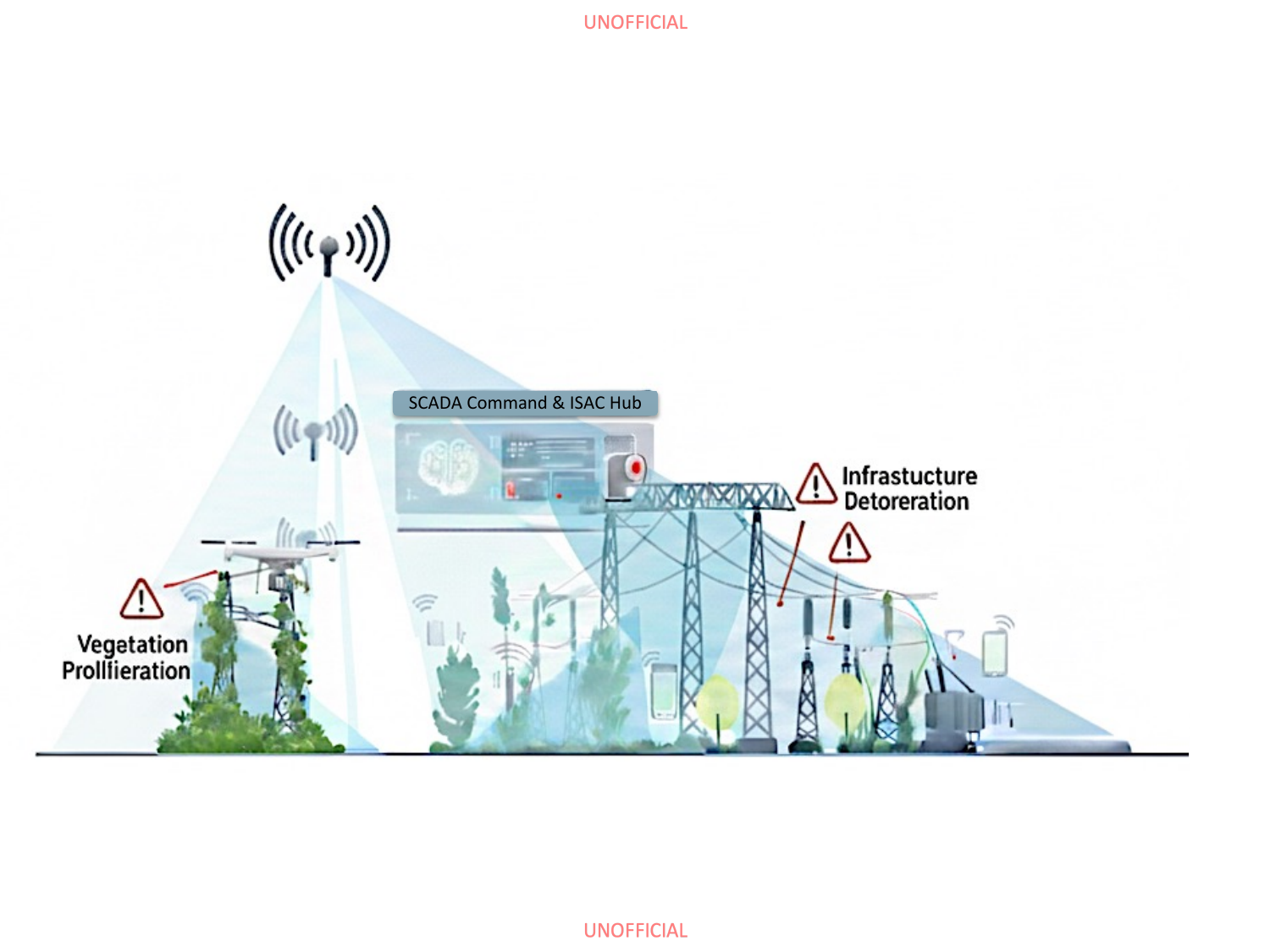} % Image for subfigure A
    \caption{}
    \label{fig:isac_grid}
\end{subfigure}
\hspace{-0.2cm} % Adds horizontal space between subfigures
\begin{subfigure}[b]{0.5\textwidth} % Adjust width as needed
    \centering
    \includegraphics[trim={0.6cm 2cm 2.8cm 2.6cm},clip, width=\textwidth]{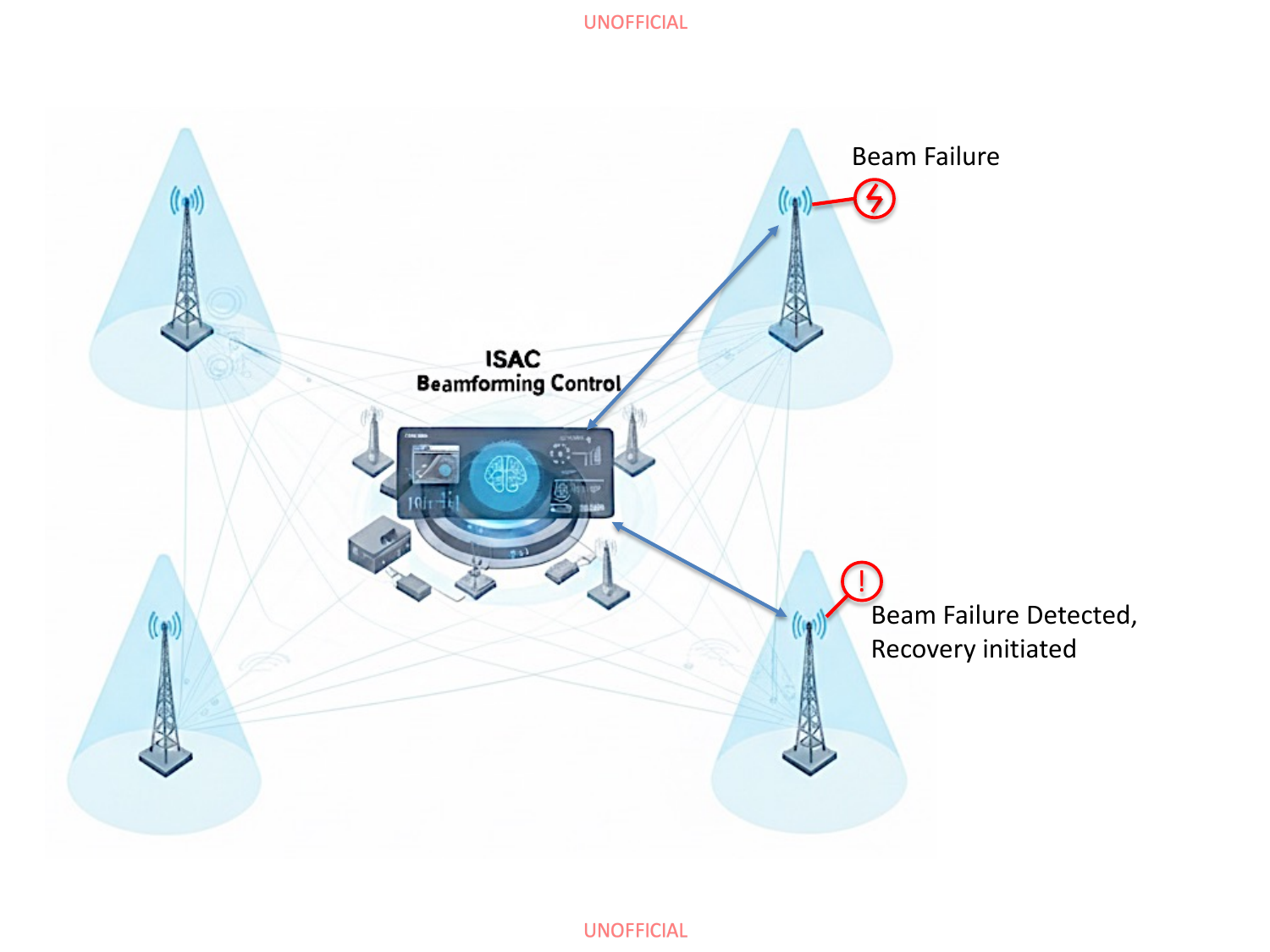} % Image for subfigure B
    \caption{}
    \label{fig:isac_com}
\end{subfigure}
\caption{An illustration of ISAC within (a) a smart grid setting, detecting infrastructure deterioration and vegetation proliferation near grid lines, and (b) a communication network, enabling predictive beamforming and recovery from beam failure.}
\label{fig:fig2}
\end{figure}

ISAC offers benefits such as unprecedented gains in spectral and hardware efficiency while reducing hardware costs~\cite{ISAC3,huawei_wp}. More importantly, it gives the network a range of powerful new capabilities that were previously limited to specialized systems. Moreover, ISAC capabilities are expected to revolutionize numerous industries and everyday life, particularly in systems such as intelligent transportation~\cite{ISAC_survey1,nokia_blog}, smart grids~\cite{smart_grid1}, smart cities~\cite{ericsson_blog}, and communication networks~\cite{huawei_wp}. For illustrations of ISAC, refer to figures Fig.~\ref{fig:isac_city} and Fig.~\ref{fig:fig2}.

Overall, this evolution transforms the wireless network into what engineers call an omnipresent sensor, forcing us to confront a new reality. While faster downloads are an easy concept to sell, environmental perception is abstract, complex, and, for many, deeply unsettling. It sounds like surveillance because, without the proper safeguards, it could be. This raises a new question: no longer just how much faster our downloads will be, but what the network will see.
%Overall, the future generation `\textbf{G}' of wireless communication networks is not just about speed; it is about giving our network eyes. Before we let it see, we must understand what it means for the network to perceive, decide what it is allowed to look at, and who is held responsible when its perception is deceived. 

For all its promise, the fusion of sensing and communication fundamentally alters the security landscape~\cite{oscar}. Security is a complicated concept, and its meaning changes when perception itself becomes the battlefield. If a network can perceive reality, that perception can be manipulated. In addition, the challenge shifts from protecting data content to safeguarding the integrity of perception itself. An adversary’s goal is no longer just to steal data, but to distort the system's perception of reality itself, leading to potentially catastrophic physical consequences. This is the weaponization of perception, and it requires a complete rethinking of what it means to be secure~\cite{PLS_SA,oscar}.

This paper examines the new reality created by ISAC, with a focus on its significant security implications. It explores the transformative opportunities this technology offers and analyzes the core challenges related to security, privacy, and trust that arise when a network can perceive the physical world. We argue that this shift renders traditional security models outdated and introduces a new range of threats that target perception itself. In response, we propose a research roadmap based on a multi-layered, defense-in-depth framework to guide future development and ensure the secure and trustworthy deployment of ISAC in the future `\textbf{G}' era, including 6G.

%====================================================
\begin{tcolorbox}[breakable, colback=blue!1!white,colframe=blue!10!black,title=Key insights]
  \begin{enumerate}[leftmargin=*]
    \item The next generation of wireless technology, such as 6G, represents more than just an upgrade; it is a fundamental shift where the network learns the ability to perceive its environment, a concept known as Integrated Sensing and Communication (ISAC).

    \item This fusion of sensing and communication fundamentally alters the security landscape, creating a new class of threats that target our perception of reality itself.

    \item Our traditional security measures are unable to defend against these new threats, which require a new focus on securing the integrity of perception.

    \item To establish trust in ISAC, proactive, defense-in-depth strategies are necessary from the outset, including the development of global standards and the integration of comprehensive safeguards and accountability measures.
\end{enumerate}
\end{tcolorbox}
%-------------------------------

%=========================================
\section{The evolution of wireless: Sensing as the next layer of the value chain}
% \begin{figure}[th]
%     \centering
%     \includegraphics[trim={1cm 3cm 1cm 4cm},clip, width=0.6\linewidth]{img1b.pdf}
%     \caption{An illustration of Integrated Sensing and Communication (ISAC) within a smart grid setting, detecting infrastructure deterioration and vegetation proliferation near grid lines.}
%     \label{fig:isac_grid}
% \end{figure}

\begin{figure}[t]
    \centering
    \includegraphics[trim={1.5cm 4.5cm 2.8cm 4.5cm},clip, width=0.8\linewidth]{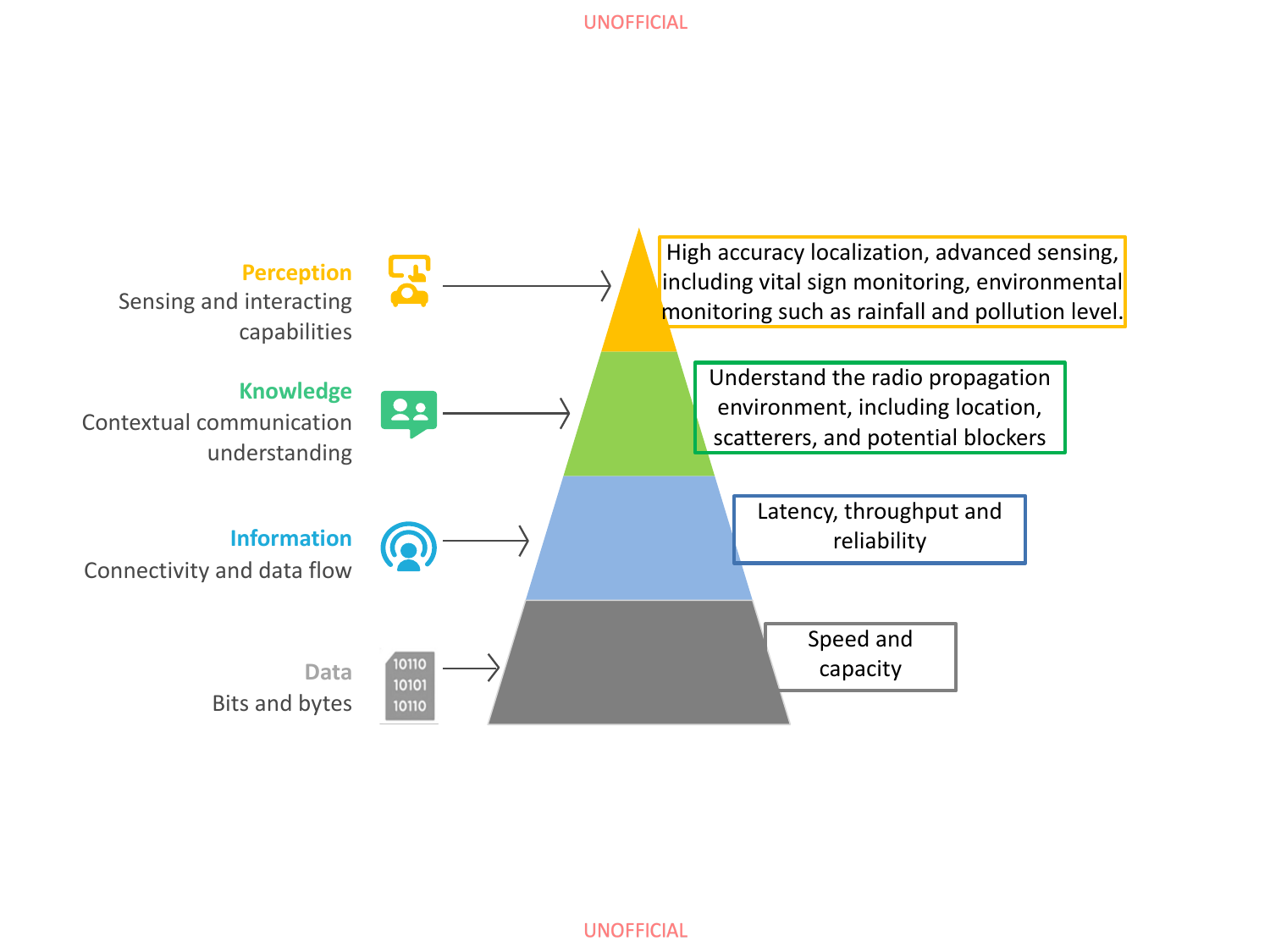}
    \caption{The evolution of wireless networks will lead to ISAC, positioning it at the top of the value chain, shifting the focus from communication to cyber-physical perception.}
    \label{fig:evolution}
\end{figure}

This evolution represents a progression along a new value chain, ending in perception. This is depicted in Fig.~\ref{fig:evolution}. The fundamental layers, Data (bits and bytes) and Information (connectivity and data flow), have traditionally been the main focus of wireless development. The network then advances to Knowledge, where it builds a contextual understanding of its radio environment~\cite{fan}. The final and most revolutionary layer is Perception, where the network gains the ability to see and interact with the physical world. This is achieved by using the same radio signals for both communication and sensing, similar to echolocation. Enabled by shifting to higher frequency bands, such as millimeter-wave (mmWave) and terahertz (THz) frequencies, and utilizing massive Multiple-Input Multiple-Output (MIMO) antenna arrays, communication signals gain radar-like capabilities~\cite{lima}. By analyzing reflected signals, the network can map its environment and determine the location, speed, and movement of objects and people~\cite{shit}.

% The unification of sensing and communication, leading to the emergence of ISAC, unlocks unprecedented gains in spectral and hardware efficiency while reducing costs. 
%
%However, its true potential lies in the new capabilities it unlocks, fundamentally changing how digital systems interact with reality. The network transitions from a passive conduit to an active participant, compelling a re-evaluation of its purpose and security requirements. 
%

More profoundly, it changes the network's role from a passive conduit to an active participant that interacts with physical reality. However, this new capability fundamentally alters the security landscape~\cite{wei}. If the network can sense, its perception can be deceived, creating vulnerabilities with direct physical consequences.
%~\cite{furqan}.

%====================================================
\section{ISAC's promise and risks: A dual-use dilemma}
%Promise
ISAC presents a classic dual-use dilemma: the same capability that provides significant benefits can also be misused for malicious or oppressive purposes. In this section, we detail both the benefits and risks associated with ISAC.  
\subsection{ISAC's promises}
To understand the stakes, we first need to recognize the world ISAC promises.

\subsubsection{Transformative opportunities}
The capabilities enabled by ISAC are set to transform numerous industries by connecting the physical and digital worlds~\cite{ericsson_wp, huawei_wp}. 
\begin{itemize}[leftmargin=*]
    \item \textbf{Smart cities and intelligent transportation systems}: ISAC provides comprehensive, real-time situational awareness across entire urban areas, applicable to public safety, such as crowd density estimation, urban planning, and environmental management~\cite{ericsson_blog}. In the intelligent transportation systems of the near future, ISAC will be the invisible guardian of our roads. Today, an autonomous car is limited by its own sensors. With ISAC, a truck will be able to perceive the child chasing a ball around a blind corner, not because its own sensors see the child, but because the network itself senses the child's presence and alerts every vehicle in the vicinity. This raises awareness of physically invisible hazards, a crucial step toward a future of cooperative safety and collision avoidance~\cite{huawei_wp,ISAC_survey1,nokia_blog,ISAC_types}.
    \item \textbf{Healthcare and augmented human sensing}: The most personal uses are in healthcare. The network's capacity to pick up subtle changes in reflected radio signals enables non-invasive tracking of vital signs~\cite{health1,health2}. Imagine a system that can monitor a sleeping infant's breathing without wires, continuously checking the stress level, or an elderly parent whose health is quietly watched by the surrounding environment itself, which can detect subtle changes in movement or behavior that might signal a health issue long before it becomes a crisis.  
    \item \textbf{Environmental monitoring and smart grids}: The network's senses extend to the broader environment as well. By analyzing how radio waves are affected by atmospheric conditions, ISAC can monitor precipitation, fog, and even pollution levels in real-time~\cite{huawei_wp}. For critical infrastructure, this means being able to detect vegetation growth near power lines before it can cause an outage, ensuring the resilience and security of the smart grid~\cite{smart_grid1}. 
    \item \textbf{New business models}: The capabilities of ISAC can also create new business models like Sensing-as-a-service (SaaS), where network operators can monetize processed sensing data through innovative means~\cite{seccure_isac1,secure_isac2}, such as providing the data to third-party clients via Application Programming Interfaces (APIs).     
\end{itemize}

% AI and ISAC
%
\subsubsection{AI/ML-powered ISAC}
Additionally, the integration of Artificial Intelligence and Machine Learning (AI/ML) enhances the capabilities of ISAC, resulting in an intelligent ISAC. This allows ISAC systems to adapt quickly to their environment and specific tasks. AI/ML brings significant improvements in several key areas~\cite{ai1,ai2,ai3}:
\begin{itemize}[leftmargin=*]
    \item \textbf{Waveform design}: Optimizing dual-purpose radio signals to balance the conflicting demands of high-speed data transmission and high-accuracy environmental sensing.
    \item \textbf{Predictive beamforming}: Directing signal beams toward where a user is expected to be, crucial for maintaining stable connections for mobile users while reducing network overhead. 
    \item \textbf{Interference management}: AI-driven interference management actively minimizes unwanted signals that can interfere with either function, ensuring clear and reliable communication. Moreover, intelligent resource allocation enables the smart assignment of network power and spectrum to maximize the system's potential.
    \item \textbf{Advanced data fusion}: Combining information from ISAC's radio sensing with data from other sensors (e.g., cameras, LiDAR) to create a single, highly robust, and unified perception of the environment.
\end{itemize}

\begin{figure}[t]
    \centering
    \includegraphics[trim={1.5cm 3.5cm 0.7cm 3cm},clip, width=0.9\linewidth]{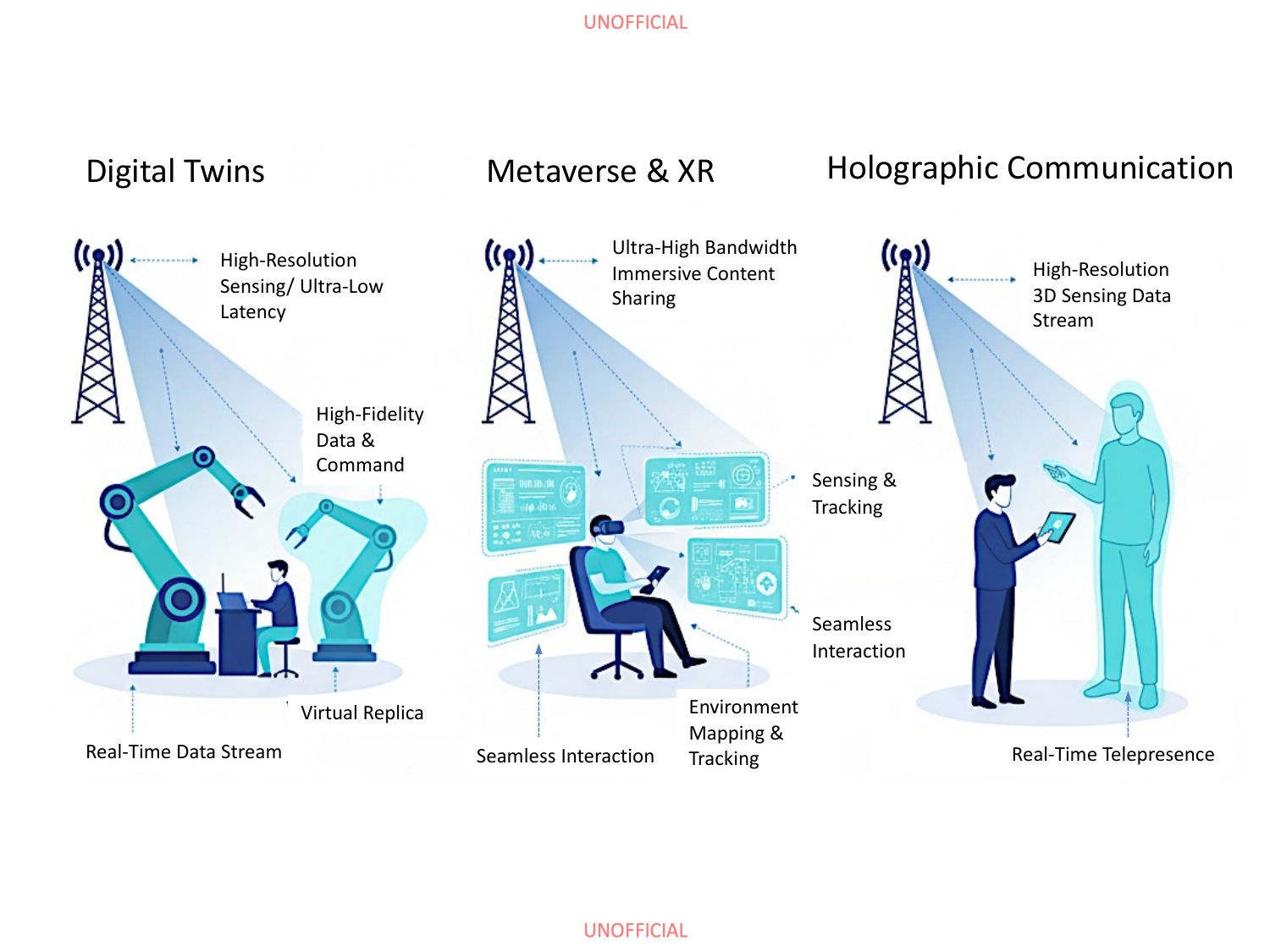}
    \caption{An illustration of ISAC as a key technology powering the next generation of immersive applications.}
    \label{fig:immersive}
\end{figure}

\subsubsection{Powering next-generation immersive applications}
ISAC is the technological foundation for a new wave of applications that merge the physical and digital worlds~\cite{vlad,zhi}. Refer to Fig.~\ref{fig:immersive} for an illustration. 
\begin{itemize}[leftmargin=*]
    \item \textbf{Digital twins}: For digital twins, which are high-fidelity virtual replicas of physical objects or systems, ISAC is essential; its continuous, high-resolution sensing provides the real-time data stream needed to keep the virtual model perfectly synchronized with its real-world counterpart, while the integrated communication link delivers these updates with ultra-low latency. 
    \item \textbf{Metaverse and extended reality (XR)}: In the metaverse and extended reality (XR), a persistent virtual universe that interacts with our physical space, ISAC's role is twofold: it delivers the ultra-high bandwidth required for streaming immersive content and, crucially, uses its sensing capabilities to map the user's physical environment and track their movements, enabling seamless and natural interaction. 
    \item \textbf{Holographic communications}: For holographic communications, the ultimate form of telepresence, ISAC not only provides the extreme bandwidth needed for real-time holographic data stream but can also perform the high-resolution 3D sensing required to generate the holographic content itself. This transforms the network from a simple data pipe into an active creator of immersive experiences.
\end{itemize}

%--------------------------------
% Peril
\subsection{ISAC's risks}
%The security challenges posed by ISAC are both technical and societal. It presents a classic dual-use dilemma: the same capability that provides significant societal benefits can also be misused for malicious or oppressive purposes. 

%====================================================
\subsubsection{Cyber-physical threat landscape}
\begin{figure}
    \centering
    \includegraphics[trim={2cm 6cm 2.5cm 5cm},clip,width=0.85\linewidth]{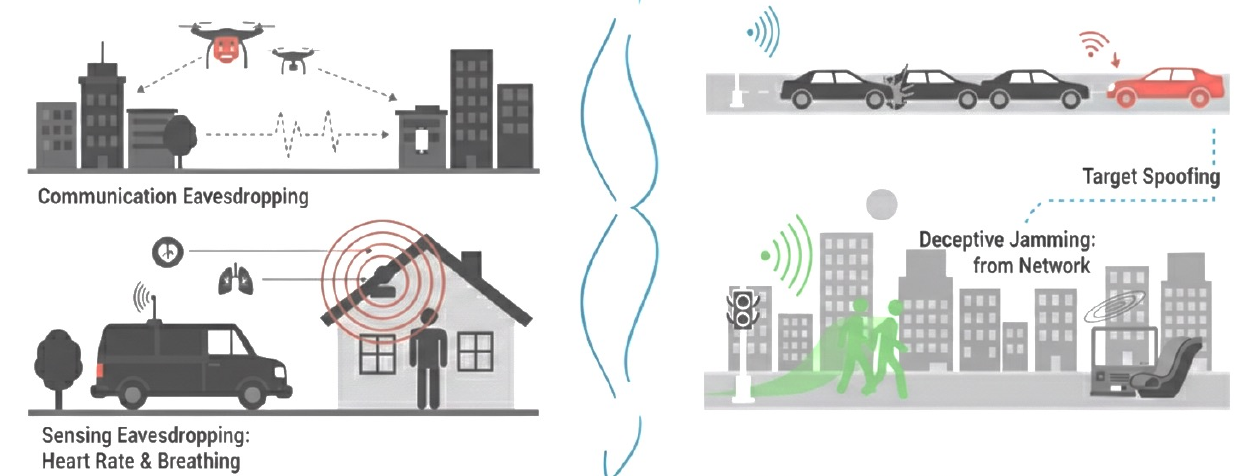}
    \caption{An illustration of threat examples in ISAC.}
    \label{fig:isac_threats}
\end{figure}

ISAC extends the digital attack surface into the physical domain, where radio waves are both perceptual tools and offensive vectors. The attacks exploiting this new vulnerability are not based on malicious code, but on the clever manipulation of sensing perception. Refer to Fig.~\ref{fig:isac_threats} for an illustration.  

\paragraph{Passive threats: Intelligence gathering}
\begin{itemize}[leftmargin=*]
    \item \textbf{Sensing eavesdropping:} The same technology that can monitor your heart rate for your doctor can be used by an adversary from a van down the street to infer your precise location, movements, and even vital signs like breathing and heart rate~\cite{deceptive_jamming}. This is a form of surveillance that can see through walls, operates without a camera, and can be conducted without your knowledge or consent, raising profound privacy concerns~\cite{Privacysecurity1}.
    \item \textbf{Communication eavesdropping:} A malicious target being tracked by an ISAC system (e.g., a hostile drone) could intercept confidential communication data embedded within the sensing waveform.
\end{itemize}

\paragraph{Active threats: Deception and disruption}
\begin{itemize}[leftmargin=*]
    \item \textbf{Target spoofing (phantom dangers):} An attacker can create phantom dangers by transmitting counterfeit radio signals that perfectly mimic the echo of a non-existent object~\cite{5Gamerica_whitepaper,target_spoofing}. Imagine an attacker generating the phantom echo of a stalled car just over a hill. An entire platoon of autonomous vehicles, perceiving an imminent threat, could slam on its brakes simultaneously, causing a catastrophic pileup. The attacker leaves no digital trace on the vehicles themselves because nothing was ever digitally compromised.
    \item \textbf{Deceptive jamming (invisible threats):} An attacker can make real threats invisible. By intercepting and altering authentic signals, they can manipulate the perceived attributes of a real target, making a pedestrian appear to be moving slower than they are, causing a vehicle's control system to misjudge the timing of an intersection, or making a legitimate vehicle vanish from the network's perception for a few critical seconds, completely undermining cooperative safety systems~\cite{RISbasedPLS,deceptive_jamming,surveysecuritysmart}.
    \item \textbf{CSI manipulation attacks:} A sophisticated attack that injects subtle interference to manipulate the Channel State Information (CSI) used by sensing systems~\cite{CSI_1}, potentially causing safety-critical failures, like an in-cabin monitor failing to detect a child in a hot car.

\end{itemize}

\begin{figure}[t]
    \centering
    \includegraphics[trim={1.6cm 2cm 0.4cm 3.5cm},clip, width=0.8\linewidth]{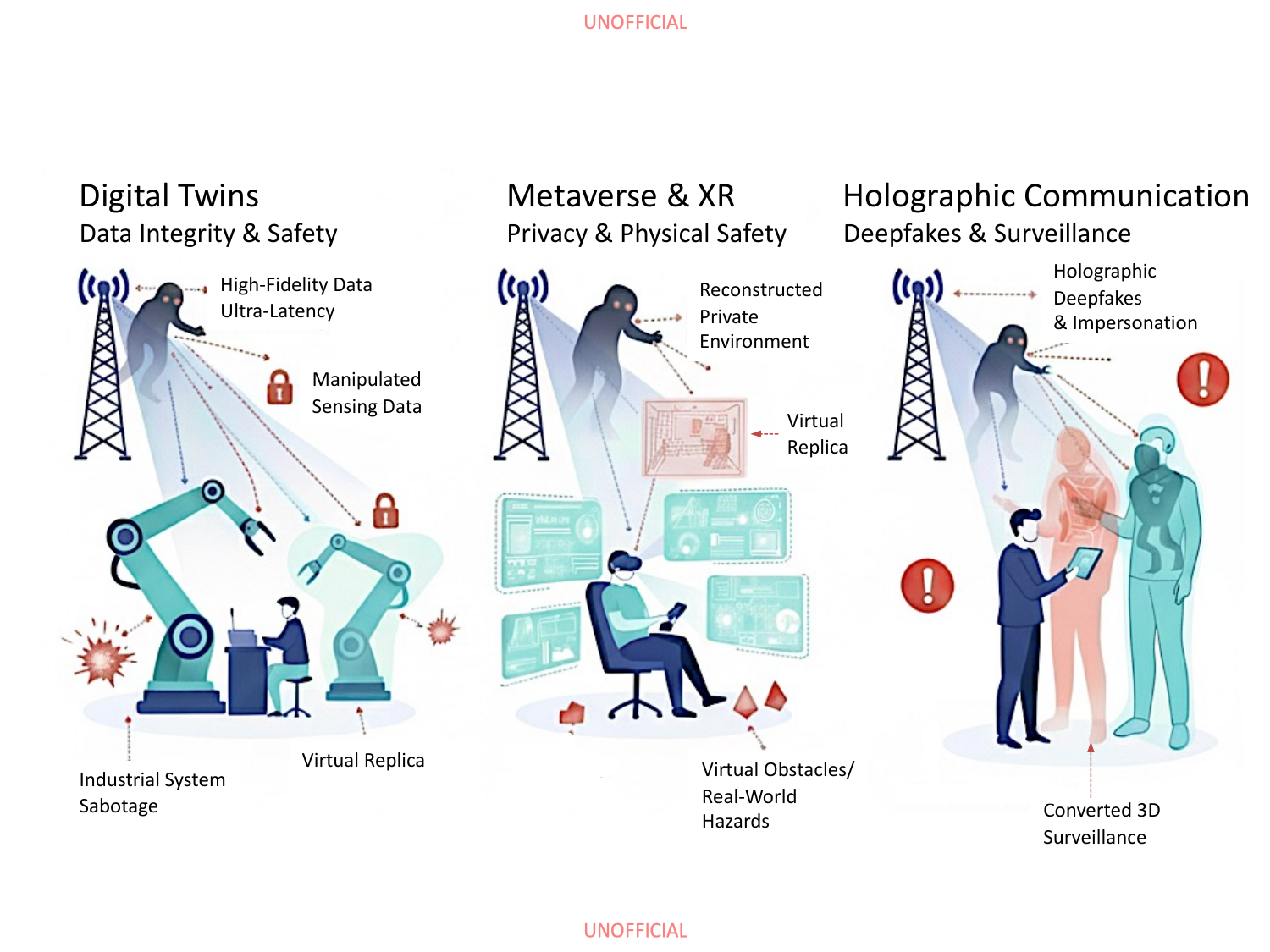}
    \caption{An illustration of unique security and privacy challenges introduced in ISAC-powered immersive applications.}
    \label{fig:immersive_security}
\end{figure}

\subsubsection{New security threats to immersive applications}
The use of ISAC to power next-generation immersive applications introduces unique security and privacy challenges by deeply intertwining data streams that represent and interact with the physical world~\cite{terry,Wang_2023}. Refer to Fig.~\ref{fig:immersive_security} for an illustration. The dual-use signal, which carries both communication data and sensing information, expands the attack surface, where compromising one function can directly affect the other, leading to new threats. 
\begin{itemize}[leftmargin=*]
    \item \textbf{Digital twins}: For digital twins, the main vulnerability lies in the integrity of the real-time synchronization link; an attacker could manipulate sensing data to feed a false reality to the virtual model, potentially causing catastrophic failures in industrial systems or leaking sensitive operational data. 
    \item \textbf{Metaverse and XR}: In the metaverse and XR, ISAC poses significant privacy risks, as an adversary could intercept sensing signals to reconstruct a user's private home, monitor their activities, and build detailed behavioral profiles. Additionally, manipulating this environmental data could pose a threat to physical safety by creating virtual obstacles or obscuring real-world hazards.
    \item \textbf{Holographic communications}: For holographic communications, ISAC's content generation capability introduces new threats, such as ``holographic eavesdropping,'' covert 3D surveillance, and the creation of three-dimensional ``deepfakes'' for malicious impersonation through alteration of the holographic data stream.
\end{itemize}

\subsubsection{``Big Brother'' scenario}
ISAC can be misused to create a potent infrastructure that, if misused, could enable a ``Big Brother'' scenario of mass surveillance and social control. A system designed to detect an unauthorized drone near a power plant can be used to track the movements of peaceful gatherings without their knowledge or consent. The technology for non-invasive healthcare monitoring could be exploited for warrantless surveillance by commercial entities. 
Furthermore, ISAC is a critical enabler for autonomous systems, providing the real-time situational awareness they need to operate effectively. While beneficial for industrial robots, this raises ethical concerns when decisions are made and applied without human control, specifically in lethal systems. 

\subsubsection{Liability crisis}
This new way of perceiving the world also creates a profound and unresolved crisis of liability. If a self-driving car has an accident because it acted on spoofed sensing data, who is at fault? The network operator who provided the flawed data? The city that used the service? The vehicle manufacturer whose systems relied on it? Our current legal frameworks rely on clear fault lines; however, in a world of shared, manipulated perceptions, those lines become increasingly blurred. This creates a legal and regulatory black hole.

Overall, ISAC raises a socio-technical challenge forcing us to ask difficult questions about the balance between security and liberty, convenience and privacy. The answer must be encoded not just in algorithms, but also align with our values, laws, and policies.

%====================================================
\section{Inadequacy of traditional defenses}
The traditional security approach centered on Confidentiality, Integrity, and Availability, the iconic CIA triad, may no longer be enough~\cite{Goll}. The challenge is no longer just protecting data content but safeguarding the integrity of perception itself. An adversary's goal is not just to steal a password or digital information, but to deceive a control system into making a disastrous decision in the physical world.

Our traditional cybersecurity defenses are no match for these attacks. Take a perimeter defense, such as a firewall, for instance. It is like a fortress wall meant to protect a digital boundary. But these attacks do not breach the perimeter; they are malicious signals sent through the air, going right around the fortress wall. When an attacker can manipulate what we see, the fortress wall is just a pretense. Another example includes signature-based detection, which is designed to identify the known fingerprints of malicious patterns or anomalies. But a spoofed radio signal has no anomalies or malicious signatures. It is a physically plausible echo whose parameters, like timing and frequency, have simply been crafted to lie. To the receiver, it looks like a legitimate reflection from the physical world. While vital for data confidentiality, encryption offers no protection if the data's content has already been compromised at the source. Encrypting a false location measurement from a spoofing attack does not make the information correct; it merely secures the transmission of a falsehood.  

Detecting such an attack requires an entirely different approach. Now we can conclude: if perception can be fooled so completely, we cannot rely on any single sensor or system. Trust in this new world cannot be taken for granted; it must be consistently earned through cross-verification and a fundamentally new, more resilient security architecture.

%====================================================
\section{A research roadmap for secure ISAC: Defense-in-depth}
Markets, fueled by a race for features, would not build this secure ecosystem on their own. The appeal of surveillance capitalism (e.g., sensing-as-a-service) and the push to cut costs will always lead to underinvestment in the strong security and privacy measures we need in the changing security landscape of ISAC. To address all aspects and build overall trust in ISAC, a defense-in-depth strategy is required, with interconnected safeguards at every level of the system. This roadmap is based on four core principles, which represent key research directions.

\begin{figure} [t]
    \centering
    \includegraphics[trim={4.8cm 4.4cm 4cm 4.8cm},clip, width=0.5\linewidth]{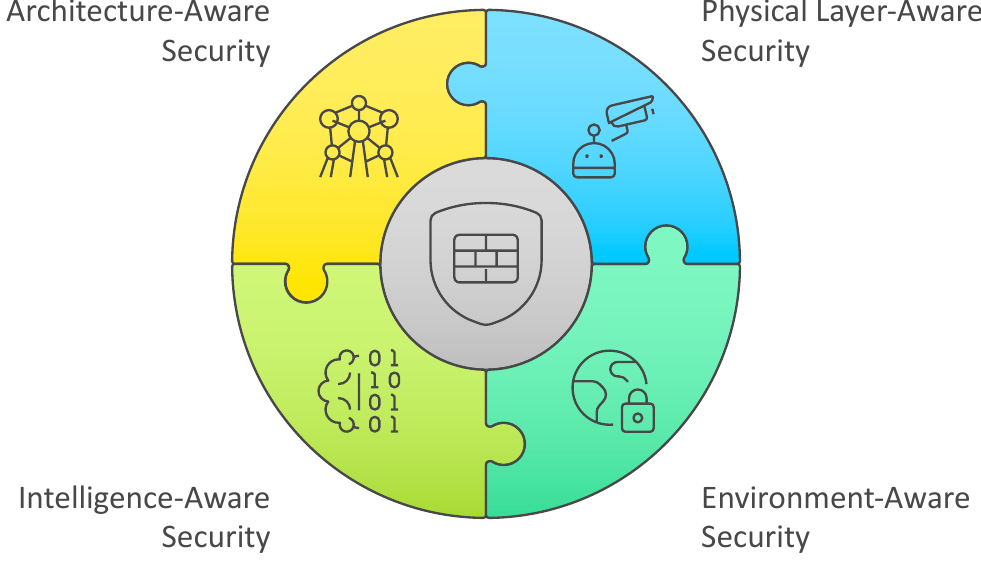}
    \caption{A multi-layered defense-in-depth framework for ISAC, where security is integrated to create a synergistic defense.}
    \label{fig:layered_defense}
\end{figure}

\begin{enumerate}[leftmargin=*]
    \item \textbf{Physical layer-aware security (smarter signals):} This principle focuses on designing the radio waves themselves to be more secure at the most fundamental level and operates directly at the physical layer~\cite{RISbasedPLS}. This includes techniques like secure beamforming with artificial noise, which directs a signal's energy toward a legitimate user while creating areas of minimal power, or nulls, in the direction of a suspected eavesdropper so that it cannot decode the message~\cite{PLS1,PLS_AN}. The sensing capability of ISAC can be used first to locate the eavesdropper, and then to jam their reception effectively, creating a synergy known as sensing-assisted security~\cite{PLS_SA}.
    \item \textbf{Environment-aware security (smarter environments):} This principle focuses on intelligently controlling and shaping radio waves within the environment itself in real-time, rather than solely at the transmitter. Using technologies like Reconfigurable Intelligent Surfaces (RIS), which are metasurfaces that can be installed on buildings, allows us to intelligently steer and shape radio waves in real-time~\cite{RISbasedPLS}. An RIS can be programmed to amplify signals for legitimate users while simultaneously creating destructive interference patterns at an attacker's location, effectively establishing dynamic secure zones~\cite{RISbasedPLS,RIS_basedISAC}.
    \item \textbf{Intelligence-aware security (smarter defenses):} Artificial intelligence/machine learning (AI/ML) is essential for building a proactive and adaptive defense. AI/ML models can be trained to monitor the perception attack, identifying threats that rule-based systems would miss. AI/ML can also coordinate a real-time, adaptive defense by optimizing beamforming and RIS configurations. However, AI/ML models can themselves be attacked (e.g., data poisoning and evasion attacks)~\cite{Privacysecurity1}. Thus, it is crucial to develop trustworthy and robust AI/ML systems for ISAC's overall security.
    \item \textbf{Architecture-aware security (smarter architecture):} The nature of ISAC demands Zero-Trust Architecture (ZTA), which operates on the principle of ``never trust, always verify". In a ZTA, every device, sensor, and user must constantly authenticate its identity and prove the integrity of its data before being granted access, no matter where it is located~\cite{zta}. This enables the elimination of the outdated concept of a trusted internal network perimeter.
\end{enumerate}

These four layers are designed to work in synergy. For example, an AI/ML-based detector (Intelligence layer) that identifies a suspicious signal can trigger the system to use sensing-assisted security (Physical layer) to locate the source. This information allows the ZTA (Architecture layer) to revoke the malicious node’s credentials while the RIS (Environment layer) directs destructive interference toward the attacker's exact location.
%====================================================
\section{Building the ecosystem: standardization, validation, and governance}
To enable overall trust in ISAC, along with the technical solutions, the unified efforts across standardization, validation, and governance must be built.
\begin{enumerate}[leftmargin=*]
    \item \textbf{Global standards for security, not just performance:} It is essential for international organizations working on ISAC, e.g., ITU, 3rd Generation Partnership Project (3GPP), and European Telecommunications Standards Institute (ETSI), to focus on security and privacy right from the start, making them core parts of their design rather than afterthoughts. Security should not be something added later; instead, it needs to be built into the very foundation of our global networks to ensure safety and trust.
    \item \textbf{Experimental validation and testbeds:} Theoretical analysis and simulations are not enough. Security-focused physical testbeds are crucial for validating defenses against realistic cyber-physical attacks, refining algorithms with real-world data, and understanding the impact of hardware limitations.
    \item \textbf{Governance, auditable systems and accountability:} We need technical and legal systems that guarantee transparency and accountability. This involves requiring the creation of immutable verifiable logs for all sensing requests, so we can identify who is sensing what and why. It also involves defining clear lines of liability that are currently absent, ensuring there is a way to hold someone accountable when perception systems fail. In addition, ISAC's ability to enable pervasive surveillance demands a new social contract to define reasonable privacy boundaries, backed by strong ethical and legal frameworks to prevent misuse.
\end{enumerate}
%======================================================
\section{Conclusion}
Integrated Sensing and Communication (ISAC) represents a monumental shift in wireless technology, offering transformative capabilities while also introducing unprecedented security challenges in the future \textbf{G} networks. The potential of this technology is far greater than what has been achieved so far, but its success depends on our ability to build a network whose perception can be trusted. This requires moving beyond traditional cybersecurity and embracing a `defense-in-depth' philosophy that deeply embeds a multi-layered defense architecture into the foundation of future generations' communication networks, including 6G. By proactively addressing the technical, ethical, and governance challenges, we can work to ensure that the future cyber-physical world is not only powerful and efficient but also secure, private, and worthy of public trust.

%====================================================
\begin{acks}
This research paper is conducted under the 6G Security Research and Development Project, as led by the Commonwealth Scientific and Industrial Research Organization (CSIRO), through funding appropriated by the Australian Government's Department of Home Affairs. This paper does not reflect any Australian Government policy position. For more information regarding this Project, please refer to \url{https://research.csiro.au/6gsecurity/}.
\end{acks}
\bibliographystyle{ACM-Reference-Format}
\bibliography{sample-base}
%====================================================

\appendix
%====================================================

\end{document}